**The emergence of power-law distributions of inter-spike intervals characterizes status epilepticus induced by pilocarpine administration in rats.**

Massimo Rizzi, Ph.D. [1]

[1] ARCEM – Associazione Italiana per la Ricerca sulle Patologie Cerebrali e del Midollo Spinale (Italian Association for the Research on Brain and Spinal Cord Diseases), Via A. Diaz 7, 20010 Vittuone (MI), Italy

Correspondence to:

Massimo Rizzi,  ARCEM – Associazione Italiana per la Ricerca sulle Patologie Cerebrali e del Midollo Spinale (Italian Association for the Research on Brain and Spinal Cord Diseases), Via A. Diaz 7, 20010 Vittuone (MI), Italy, E-mail: massimo.rizzi@arcem.it

**Summary**

*Objective.* To ascertain the existence of power-law distributions of inter-spike intervals (ISI) occurring during the progression of status epilepticus (SE), so that the emergence of critical states could be reasonably hypothesized as being part of the intrinsic nature of the SE.

*Methods.* Status epilepticus was induced by pilocarpine administration in post-natal 21-day rats (n=8). For each animal, 24 hours of EEG from the onset of the SE were analyzed according to the analytical procedure suggested by Clauset et al. (2009) which combines maximum-likelihood fitting methods with goodness-of-fit tests based on the Kolmogorov-Smirnov statistics and likelihood ratios. The analytical procedure was implemented by the freely available R software package 'poweRlaw'. Time of calculations was considerably shorten by the exploitation of High-Throughput Computing technology, a.k.a. Grid Computing technology.

*Results.* The progression of the SE is characterized by the emergence of power-law correlations of ISI whose likelihood of occurrence increases the more the time from the onset of the SE elapses. Log-normal distribution of ISI is however widely represented. Additionally, undetermined distributions of ISI are represented as well, although confined within a restricted temporal window. The final stage of SE appears dominated only by power-law and log-normal distributions of ISI.

*Significance.* The emergence of power-law correlations of ISI concretely supports the concept of the occurrence of critical states during the progression of SE. It is reasonably speculated, as a working hypothesis, that the occurrence of power-law distributions of ISI within the early stages of the SE could be a hallmark of the establishment of the route to epileptogenesis.

**Highlights**

- We ascertained the emergence of power-law correlations during the status epilepticus.
- Log-normal correlations are however widely represented.
- Undetermined correlations also occur, but limited to the first 14-18 hours.
- Status epilepticus could evolve through the occurrence of criticalities.

# 1. Introduction

A pathogenic phenomenon strictly related to the occurrence of epilepsy is the status epilepticus (SE), a condition defined in humans as a seizure that persists for at least 5-10 minutes or is repeated frequently enough not to have a resolution of contiguity (Lowenstein and Alldredge, 1998). SE is potentially life-threatening and is positively correlated with the risk of developing epilepsy during the lifetime (Hesdorffer et al., 1998). However, although in humans the onset of the SE represents a clinical emergency, the same insult is widely exploited in the basic research, since SE is commonly induced in rodents in order to develop epilepsy intentionally. This experimental paradigm allows studies spanning from the identification of possible markers of epileptogenesis to the evaluation of potential new therapeutic treatments for the prevention of epileptogenesis or as disease-modifying drugs (Löscher and Brandt, 2010; Pitkanen, 2010).

Despite the experimentally-induced SE may offer a unique occasion to shed light on the mechanisms which induce epileptogenesis, the nature of phenomena elicited during the progression of the SE is poorly understood.

The experimentally-induced SE is electrographically characterized by a progression of epileptiform discharges and seizures, interspersed each others, giving rise to an intense spiking activity lasting several hours, showing the highest spiking rates to occur within the early stages of the SE (1-3 hours from the onset), followed by a progressive decrease the more the time elapses (Treiman et al., 1990; Pitkänen et al., 2005; Lehmkuhle et al., 2009). However, as it is well known to experimentalists, spiking rates can show broad ranges of variation, even within narrow observational time windows. Although, on the one hand, such fluctuations could be uncorrelated, on the other hand, they could suggest the spiking activity as being the expression of nonlinear phenomena occurring during the progression of the SE. Among nonlinear phenomena of interest in the context of brain dynamics, the establishment of power-law correlations of observables deserves attention since they emerge in association with the occurrence of a critical state, which is a condition denoting a border between qualitatively different types of behavior that a complex nonlinear system can show. In this state, the magnitude of an observable can vary according to a power-law distribution, hence, that observable cannot be depicted by a representative value, as the mean. In such condition, the observable is 'scale-free' and it is considered as the hallmark that a new behavior can emerge in the temporal evolution of the system.

Power-law correlations have been reported to occur in physiological conditions (Bak et al., 1988; Jensen, 1998; Hesse and Gross, 2014) as well as in pathological contexts, as epilepsy. Indeed, such peculiar behavior was shown to emerge in epilepsy related phenomena such as epileptiform discharges associated to the so-called neuronal avalanches (Beggs and Plenz, 2003, 2004; Shew et al., 2009; Benayoun et al., 2010), the activity of neuronal networks involved in seizure generation (Bak et al., 1988; Worrell et al., 2002; Monto et al., 2007) and inter-seizure intervals (Osorio et al., 2009). Consistently, deviations from power-law behavior were shown to occur during seizure progression (Meisel et al., 2012).

Therefore, the emergence of power-law behavior of observables in epilepsy related phenomena may be actually associated with the occurrence of critical states, which appear as an intrinsic property of the epileptic brain and supports the notion of epilepsy as a complex dynamical disease (Beggs & Plenz, 2003, 2004; Osorio et al., 2009, 2010; Milton, 2012).

It is worth considering, however, that the great majority of the aforementioned evidence derives from studies accomplished in a context of overt pathology. Conversely, the SE represents a circumscribed insult which may or may not lead to the development of epilepsy, thus depicting a considerably different condition from the overt pathology and, in principle, the possible emergence of power-law correlations during the progression of the SE could have a different  functional significance from that (yet unknown) of the occurrence of the same phenomena in the context of epilepsy.

Nevertheless, although interesting, no studies have to date investigated the emergence of power-law correlations during the progression of the SE. Therefore, we tested this possibility by the application of an appropriate mathematical procedure (Clauset et al., 2009) to datasets of inter-spike intervals measured for 24 hours from the onset of the SE.

## 2. Materials and methods

### 2.1 Induction of status epilepticus, EEG recordings and measurement of inter-spike intervals

All procedures involving experimental animals were accomplished according to the principles set out in the following laws, regulations, and policies governing the care and use of laboratory animals: Italian Governing Law (D.lgs 26/2014; Authorisation n.19/2008-A issued March 6, 2008 by Ministry of Health); Mario Negri Institutional Regulations and Policies providing internal authorisation for persons conducting animal experiments (Quality Management System Certificate –UNI EN ISO 9001:2008 – Reg. N° 6121); the NIH Guide for the Care and Use of Laboratory Animals (2011 edition) and EU directives and guidelines (EEC Council Directive 2010/63/UE). The Statement of Compliance (Assurance) with the Public Health Service (PHS) Policy on Human Care and Use of Laboratory Animals has been recently reviewed (9/9/2014) and will expire on September 30, 2019 (Animal Welfare Assurance #A5023-01).

EEG recordings of SEs considered in this study were from 8 rats at postnatal day 21 (PN21, being PN0 defined as the day of birth) which underwent a protocol of surgical implantation of a depth electrode in the dorsal hippocampus (Marcon et al., 2009). For each rat, SE was induced according to the procedure described elsewhere (Marcon et al., 2009). Briefly, at PN20, rats were intraperitoneally injected with lithium chloride (3.36 mg/kg; Merck Sharp and Dohme, Rome, Italy) and 18–20 h later, at PN21, they were subcutaneously injected with pilocarpine (60 mg/kg; Sigma-Aldrich, St. Louis, MO, USA). Status epilepticus developed in all animals after $11.5 \pm 0.3$ minutes from the administration of pilocarpine. After 48 hours from the onset of the status epilepticus, animals were euthanized.

EEG signals were filtered (cut-off frequency 0.3 Hz - 70 Hz, notch filter 50 Hz), digitized at 256 sample per second with 16 bits precision and acquired by the computer-based system LabChart v. 7.3.7 (ADInstrument, Australia). The onset of the SE was defined as the appearance of sustained high-frequency (> 8 Hz) discharges. For each EEG recording, we considered as a single spike any event lasting >20 ms and < 200 ms and amplitude 3 times the standard deviation above and below the mean of an EEG epoch selected during the final stages of the SE, devoid of prominent spiking activity. For each animal, 24 hours of EEG recording from the onset of the SE were segmented in non-overlapping time windows of 1 hour duration. For each time window, spikes counting and measurement of related inter-spike intervals were performed by the software Clampfit v. 10.0 (Molecular Devices, Silicon Valley, CA, USA).

### 2.2 Mathematical analysis and computational resources

For each non-overlapping time window of 1 hour duration, the related inter-spike intervals (ISI) dataset was analyzed according to Clauset and colleagues (Clauset et al. 2009), who introduced a set of statistically principled methods for fitting and testing the power-law hypothesis for continuous or discrete datasets. Their approach combines maximum-likelihood fitting methods with goodness-of-fit tests based on the Kolmogorov-Smirnov (KS) statistics (Reiss and Thomas, 2007; Drees and Kaufmann, 1998) and likelihood ratios (Vuong, 1989), the latter for testing the statistical significance of the power-law hypothesis against alternate plausible distributions. Indeed, since power-law like distributions of empirical data could be due to uncorrelated events as well as to different heavy-tailed distributions, we also tested these possibilities. Accordingly, the power-law hypothesis for each ISI dataset was challenged by i) the Poisson distribution, for testing the presence of uncorrelated events, and ii) the exponential and the log-normal distributions, for testing the presence of alternate heavy-tailed distributions. It is relevant to notice that power-law distributions and log-normal distributions often compete for modeling the same dataset of measurements, especially in the nervous system (Buzsáki & Mizuseki, 2014, and references therein). This is not surprising since power-law and log-normal distributions have similar generative models which can be tuned by minor variations leading to one or the other distribution (Mitzenmacher, 2003).

Calculations were performed by the software package 'poweRlaw' (Gillespie, 2014). Computational constraints, mainly due to the implementation of the bootstrapping procedure (n=5000), were removed by High-Throughput-Computing technology, a.k.a. Grid Computing Technology (Barbera et al., 2011).

# 3. Results

Temporal profiles of spiking activity in our experimental conditions were consistent with those reported by other investigators, as shown in figure 1. Indeed, for each individual animal, the highest percentage of spikes (as respect to the total number of spikes counted over 24 hours) occurs within the first 3 hours from the onset of the SE (figure 1, panel A). This percentage quickly decreases in the following temporal windows. The general pattern of spiking activity for grouped animals is depicted in figure 1, panel B.

The evaluation of ISI distributions by heavy-tailed models shows that power-law correlations of ISI emerge during the progression of SE. The existence of power-law correlations of ISI was challenged by also considering log-normal, exponential and Poisson models of distributions for the same datasets. Table 1 concisely reports for each experimental animal the temporal profile of best-fit models of ISI distributions as evinced by the application of the analytical method according to Clauset et al. (2009, full details in Supplementary material). Exponential and Poisson distributions of ISI were never good models as compared to power-law for any dataset analyzed and the only competing model was the log-normal distribution. However, a relevant percentage of datasets was not compliant to any model of heavy-tailed distribution considered in this study, hence these ISI datasets were labeled as 'undetermined' distribution. Table 2 helps readers to quickly identify the timing of occurrence of specific ISI distributions, by the same color-code used in table 1. One easily notices that log-normal distributions of ISI are widely represented over 24 hours of SE, whereas 'undetermined' distributions seem not to occur in final stages of the SE, where only power-law and log-normal distributions appear as plausible models for ISI (tab. 1-2, Supplementary material).

**4. Discussion**

Our investigation on temporal correlations embedded in spiking activity has allowed to shed light on the nature of the ongoing dynamics during the progression of the SE. Indeed, at least three different distributions of ISI can be distinguished, two of which are clearly delineated, i.e., log-normal and power-law, whereas the third distribution, although undetermined, shows a peculiar timing of occurrence, since it fades away in long-term stages of the SE, which are apparently characterized only by power-law and log-normal distributions of ISI. Although further investigations are needed for a definitive word, it is worth considering that the emergence of power-law distributions of ISI concretely supports the hypothesis of the occurrence of critical states during the progression of the SE, thus assimilating this pro-epileptogenic insult to a complex phenomenon evolving through the occurrence of criticalities.

Although our results are intriguing, we are nonetheless aware that we cannot draw conclusions on the significance of the occurrence of power-law, log-normal and undetermined distributions of ISI and their functional role in the context of the SE. Accordingly, in the following, we limit to sketch out two hypotheses, leaving our empirical findings ' as they are' to the attention of interested readers as reference for future investigations.

As a first hypothesis, it is worth noticing that the percentage of animals experiencing the emergence of power-law distributions of ISI within the early stages of the SE (2-3 hours from the onset) is closely similar to that of animals which are expected to develop epilepsy according to our experimental protocol for age-matched rats, i.e., approximately 60% (Dubé et al., 2001; Marcon et al., 2009). Since this temporal window is known as being crucial for the development of epileptogenesis (Jones et al., 2002; Löscher and Brandt, 2010; Pitkanen, 2010), our findings could suggest that the emergence of power-law distribution of ISI in the earliest stages of the SE could characterize those animals that will develop epilepsy, whereas the occurrence of power-law distributions of ISI in later temporal windows could characterize those animals for which epilepsy will not develop and/or those with mild forms of epilepsy. From this point of view, the early emergence of power-law distributions of ISI could reflect the severity of ongoing pro-epileptogenic mechanisms elicited by the SE, so that the more anticipated the emergence of such distributions, the higher the likelihood of induction of epileptogenesis.

As a conclusive consideration/hypothesis, this study confirms that log-normal distribution of events is widely represented in the nervous system (for an interesting review on log-normal distributed events in the brain see Buzsáki and Mizuseki, 2014, and references therein), also in a pathogenic context such as the SE, maybe suggesting the SE as being the macroscopic manifestation of phenomena which usually occur at more restricted spatio-temporal scales, maybe similarly to what occurs for seizures (Stead et al., 2010). From a more mechanistic perspective, the pervasive log-normal distribution of ISI could be consistent with the synchronous discharging of different metastable (i.e., short-lived) neural networks which conceivably emerge during the progression of the SE. Indeed, investigations on the progression of the ictal activity by graph theory have shown how the electrographic temporal evolution of a seizure could be ascribed to the coalescence/fragmentation of different neuronal networks which emerge transiently (Kramer et al., 2010) and which may show small-world structures (Ponten et al., 2007), thus contributing to sustain the ongoing ictal activity (Netoff et al., 2004). Accordingly, SE could reflect a similar situation. From this point of view, the 'disappearing' of undetermined ISI distributions beyond the 14[th]-18[th] hour from the onset of the SE (depending on the animal involved) could represent the actual duration of the stages of the SE during which neuronal networks undergo substantial rearrangements. These dynamics could be the expression of transient trajectories in a context of unstable/metastable dynamical landscapes (Foss et al., 1997; Milton, 2012).

We are reasonably confident that our findings will stimulate the interest of experimentalists and promote investigations aimed to the ascertainment of the relationship, if any, between the emergence of power-law distribution of ISI and epileptogenesis, and how the occurrence of this correlation could be modulated by pharmacological interventions. It cannot be excluded that meaningful parameters useful from this perspective could be the relative percentages of occurrence of power-law, log-normal and undetermined distributions of ISI and how these are modulated by therapeutics. Scaling parameters and number of data points fitted by power-law models as well as means and standard deviations of log-normal distributions may help quantifying such effects. As a corollary derived from this future investigation, from the perspective of parameterization of the experimentally-induced SE, the emergence of power-law relationships of ISI could represent the appropriate analytical tool to monitor the consistency of the pro-epileptogenic insult elicited by proconvulsant agents among experimental groups.

**References**

Bak P, Tang C, Wiesenfeld K (1988) Self-organized criticality. Phys Rev A 38(1): 364-374.

Barbera R, La Rocca G, Rizzi M (2011) Grid Computing Technology and the Recurrence Quantification Analysis to Predict Seizure Occurrence in Patients Affected by Drug-Resistant Epilepsy. In Lin SC, Yen E (eds) Data Driven e-Science. Springer New York. pp. 493-506.

Beggs JM, Plenz D (2003) Neuronal avalanches in neocortical circuits. J Neurosci 23(35): 11167-11177.

Beggs JM, Plenz D (2004) Neuronal avalanches are diverse and precise activity patterns that are stable for many hours in cortical slice cultures. J Neurosci 24(22): 5216-5229.

Benayoun M, Kohrman M, Cowan J, van Drongelen W (2010) EEG, temporal correlations, and avalanches. J Clin Neurophysiol 27(6): 458-464.

Buzsáki G, Mizuseki K (2014) The log-dynamic brain: how skewed distributions affect network operations. Nat Rev Neurosci 15: 264–278.

Clauset A, Shalizi CR, Newman ME (2009) Power-law distributions in empirical data. SIAM review 51(4): 661-703.

Drees H, Kaufmann E (1998) Selecting the optimal sample fraction in univariate extreme value estimation. Stochastic Process Appl 75: 149–172.

Dubé C, Boyet S, Marescaux C, Nehlig A (2001). Relationship between neuronal loss and interictal glucose metabolism during the chronic phase of the lithium-pilocarpine model of epilepsy in the immature and adult rat. Exp Neurol 167(2): 227-241.

Foss J, Eurich CW, Milton J, Ohira T (1997) Noise, multistability and long-tailed interspike interval (ISI) histograms. Bull Am Phys Soc 42:781.

Gillespie CS (2014) Fitting heavy tailed distributions: the poweRlaw package. R package version 0.20.5.

Hesdorffer DC, Logroscino G, Cascino G, Annegers JF, Hauser WA (1998) Risk of unprovoked seizure after acute symptomatic seizure: effect of status epilepticus. Ann Neurol 44: 908–912.

Hesse J, Gross T (2014) Self-organized criticality as a fundamental property of neural systems. Front Syst Neurosci: 23 September 2014; doi: 10.3389/fnsys.2014.00166.

Jensen HJ (1998) Self-organized criticality: emergent complex behavior in physical and biological systems (Vol. 10). Cambridge university press.

Jones DM, Esmaeil N, Maren S, Macdonald RL (2002). Characterization of pharmacoresistance to benzodiazepines in the rat Li-pilocarpine model of status epilepticus. Epilepsy Res 50(3): 301-312.

Kramer MA, Eden UT, Kolaczyk ED, Zepeda R, Eskandar EN, Cash SS (2010) Coalescence and fragmentation of cortical networks during focal seizures. J Neurosci 30(30): 10076-10085.

Lehmkuhle MJ, Thomson KE, Scheerlinck P, Pouliot W, Greger B, Dudek FE (2009). A simple quantitative method for analyzing electrographic status epilepticus in rats. J Neurophysiol 101(3): 1660-1670.

Löscher W, Brandt C (2010) Prevention or modification of epileptogenesis after brain insults: experimental approaches and translational research. Pharmacol Rev 62(4): 668-700.

Lowenstein DH, Alldredge BK (1998) Status epilepticus. *N Engl J Med* 338(14): 970-976.

Marcon J, Gagliardi B, Balosso S, Maroso M, Noé F, Morin, Lerner-Natoli M, Vezzani A, Ravizza T (2009) Age-dependent vascular changes induced by status epilepticus in rat forebrain: implications for epileptogenesis. Neurobiol Dis 34(1): 121-132.

Meisel C, Storch A, Hallmeyer-Elgner S, Bullmore E, Gross T (2012) Failure of adaptive self-organized criticality during epileptic seizure attacks. PLoS computational biology 8(1): e1002312.

Milton JG (2012) Neuronal avalanches, epileptic quakes and other transient forms of neurodynamics. Eur J Neurosci 36(2): 2156-2163.

Mitzenmacher M (2003) A brief history of generative models for power law and lognormal distributions. Internet Math 1: 226–251.

Monto S, Vanhatalo S, Holmes MD, Palva JM (2007) Epileptogenic neocortical networks are revealed by abnormal temporal dynamics in seizure-free subdural EEG. Cereb Cortex 17(6): 1386-1393.

Netoff TJ, Clewley R, Arno S, Keck T, White AJ (2004) Epilepsy in small-world networks. J Neurosci 24 (37): 8075–8083

Osorio I, Frei MG, Sornette D, Milton J (2009) Pharmaco-resistant seizures: self-triggering capacity, scale-free properties and predictability?. Eur J Neurosci 30(8): 1554-1558.

Osorio I, Frei MG, Sornette D, Milton J, Lai YC (2010) Epileptic seizures: Quakes of the brain?. Phys Rev E Stat Nonlin Soft Matter Phys 82(2): 021919.

Pitkänen A, Kharatishvili I, Narkilahti S, Lukasiuk K, Nissinen J (2005) Administration of diazepam during status epilepticus reduces development and severity of epilepsy in rat. Epilepsy Res 63(1): 27-42.

Pitkänen A (2010). Therapeutic approaches to epileptogenesis—hope on the horizon. Epilepsia, 51(s3): 2-17.

Ponten SC, Bartolomei F, Stam CJ (2007) Small-world networks and epilepsy: graph theoretical analysis of intracerebrally recorded mesial temporal lobe seizures. Clin Neurophysiol 118(4): 918-927.

Reiss RD, Thomas M (2007) Statistical Analysis of Extreme Values with Applications to Insurance, Finance, Hydrology and Other Fields, 3rd ed. Birkhäuser, Basel.

Shew WL, Yang H, Petermann T, Roy R, Plenz D (2009) Neuronal avalanches imply maximum dynamic range in cortical networks at criticality. J Neurosci 29(49): 15595-15600.

Stead M, Bower M, Brinkmann BH, Lee K, Marsh WR, Meyer FB, Litt B, Van Gompel J, Worrell GA (2010) Microseizures and the spatiotemporal scales of human partial epilepsy. Brain, awq190.

Treiman DM, Walton NY, Kendrick C (1990) A progressive sequence of electroencephalographic changes during generalized convulsive status epilepticus. Epilepsy Res *5*(1): 49-60.

Vuong QH (1989) Likelihood ratio tests for model selection and nonnested hypotheses. Econometrica 57: 307–333.

Worrell GA, Cranstoun SD, Echauz J, Litt B (2002) Evidence for self-organized criticality in human epileptic hippocampus. Neuroreport 13(16): 2017-2021.

**Captions**

**Fig.1. Temporal profiles of spikes counting**

Temporal profile of spikes counting for each individual animal (panel A) and for all animals grouped in order to highlight the general pattern of progression of SE when parameterized by spiking activity (panel B). Data are expressed as percentage of number of spikes respect to the total amount of spikes counted during 24 hours of SE.

**Tab.1. Temporal profile of the occurrence of tested heavy-tailed distributions of ISI during 24 hours of progression of SE**

Temporal profile of the occurrence of tested heavy-tailed distributions of ISI during 24 hours of progression of SE induced in PN21 rats (n=8) by the i.p. administration of pilocarpine. Starting from the onset of the SE, ISI were measured in successive non-overlapping 24 temporal windows of 1 hour duration.

For each temporal window analyzed, the favored distribution of ISI is reported by the following codes:

- *power-law*: color-code=yellow/pink; letter-code=p;
- *log-normal*: color-code=green; letter-code=ln;
- *undetermined*: color-code=grey; letter-code=u.

Exponential and Poisson distributions were never plausible models to fit to ISI datasets, see Supplementary material for detailed statistics.

This table also reports the statistical plausibility, denoted as *p (bootstrap)*, for power-law model. According to Clauset et al. (2009), the power-law hypothesis as a fitting model of ISI must be rejected for *p (bootstrap)*$< 0.1$. Similarly to Clauset and colleagues, when the power-law hypothesis is statistically significant, we also give an indication (denoted as *plausibility*) of the goodness of such plausibility. Criteria for ranking the goodness of plausibility are reported in Supplementary material. For power-law hypothesis classified as 'likely' the color-code is pink.

**Tab.2. Timing of occurrence of heavy-tailed distributions of ISI, depicted individually**

To better highlight the timing of occurrence of heavy-tailed distributions of ISI reported in table 1, their respective time-course are depicted individually. Distributions are represented by the same color-code used in table 1.

**Supplementary material**. **Tests for statistical plausibility of the power-law model of distribution of ISI vs. log-normal, exponential and Poisson competing models**

Tests for statistical plausibility of the power-law model of distribution of ISI vs. log-normal, exponential and Poisson competing models during 24 hours of progression of SE induced in PN21 rats (n=8) by the i.p. administration of pilocarpine. Starting from the onset of the SE, ISI were measured in successive non-overlapping 24 temporal windows of 1 hour duration, so that for each experimental animal 24 hours of SE were analyzed.

For each competing distribution, the logarithm of likelihood ratio ($LR$) and the relative $p$ value are reported. The $LR$ denotes which distribution between power-law and the competing distribution has the higher probability of fitting the data. Being the $LR$ the logarithm of a ratio, its sign determines which distribution is favored. Although in principle, for negative values, the competing distribution is favored, it cannot be ruled out that the sign may be caused by a chance fluctuation, especially when the $LR$ value is close to zero. Therefore, a $p$ value (two-sided) was introduced (Vuong, 1998) in order to evaluate the statistical significance of the sign of the $LR$. The sign is statistically significant if $p < 0.05$ (Vuong, 1998). Therefore, the general condition for which a competing distribution is favored over the power-law distribution requires both a negative $LR$ and the corresponding $p < 0.05$. In this study, each dataset underwent a bootstrapping procedure (n=5000) for testing power-law hypothesis, even when competing distributions appeared to clearly favor the power-law model according to their respective $LR$ and $p$ values. The p value calculated by the bootstrapping procedure is reported in the table as $p$ (bootstrap). Exponential and Poisson distributions were never plausible models for datasets of ISI and the only plausible competing distribution as respect to the power-law hypothesis was the log-normal. However, one notices that it may sometime occurs that even when $LR$ and $p$ values of competing distributions favor power-law as best-fit model for an ISI dataset, not necessarily the power-law distribution is actually the best-fit for that dataset. Indeed, according to Clauset et al. (2009), for $p$ (bootstrap) $< 0.1$ the power-law hypothesis must be rejected. Generally, datasets for which none of the tested heavy-tailed distributions represented a plausible model were labeled as 'undetermined' if:

- $p$ (bootstrap) $< 0.1$ and $LR$ of one or more competing distributions with definitive positive value ($p < 0.05$);
- $p$ (bootstrap) $< 0.1$ and $LR$ of one or more competing distributions with uncertain positive value ($p \geq 0.05$);
- $p$ (bootstrap) $< 0.1$ and $LR$ of one or more competing distributions with uncertain negative value ($p \geq 0.05$).

When $p$ (bootstrap) $\geq 0.1$, hence power-law hypothesis was clearly favored over competing distributions, we report our judgments of the goodness of plausibility of such model for that dataset of ISI, similarly to Clauset and colleagues. Our judgments keep in consideration that the minimum number of data points (denoted in the table as $n\_tail$) fitted by the power-law model was suggested to be at least $n\_tail \cong 100$ for a reliable discrimination of power-law model over competing distributions. Specifically, the goodness of plausibility of power-law hypothesis was ranked:

- moderate, if $p$ (bootstrap) $\geq 0.1$ and $LR$ for all competing distributions with uncertain negative value ($p \geq 0.05$);
- good, if $p$ (bootstrap) $\geq 0.1$ and LR for all competing distributions with definitive positive value ($p < 0.05$);
- likely, if $p$ (bootstrap) $\cong 0.1$ or $p$ (bootstrap) $\geq 0.1$ but $n\_tail < 100$.

For each plausible power-law fit, the scaling parameter, *alpha ± SD*, and the number of data points fitted by the power-law model, *n_tail*, are reported.

Log-normal distribution was the only competing heavy-tailed model frequently favored during the progression of the SE, for which LR $< 0$ and $p < 0.05$.

For readers' convenience, we represented favored distributions by the same color-code and letter-code used in table 1 of the manuscript, as follows:

- *power-law*: color-code = yellow/pink; letter-code = p;
- *log-normal*: color-code = green; letter-code = ln;
- *undetermined*: color-code = grey; letter-code = u.

**Figure 1**

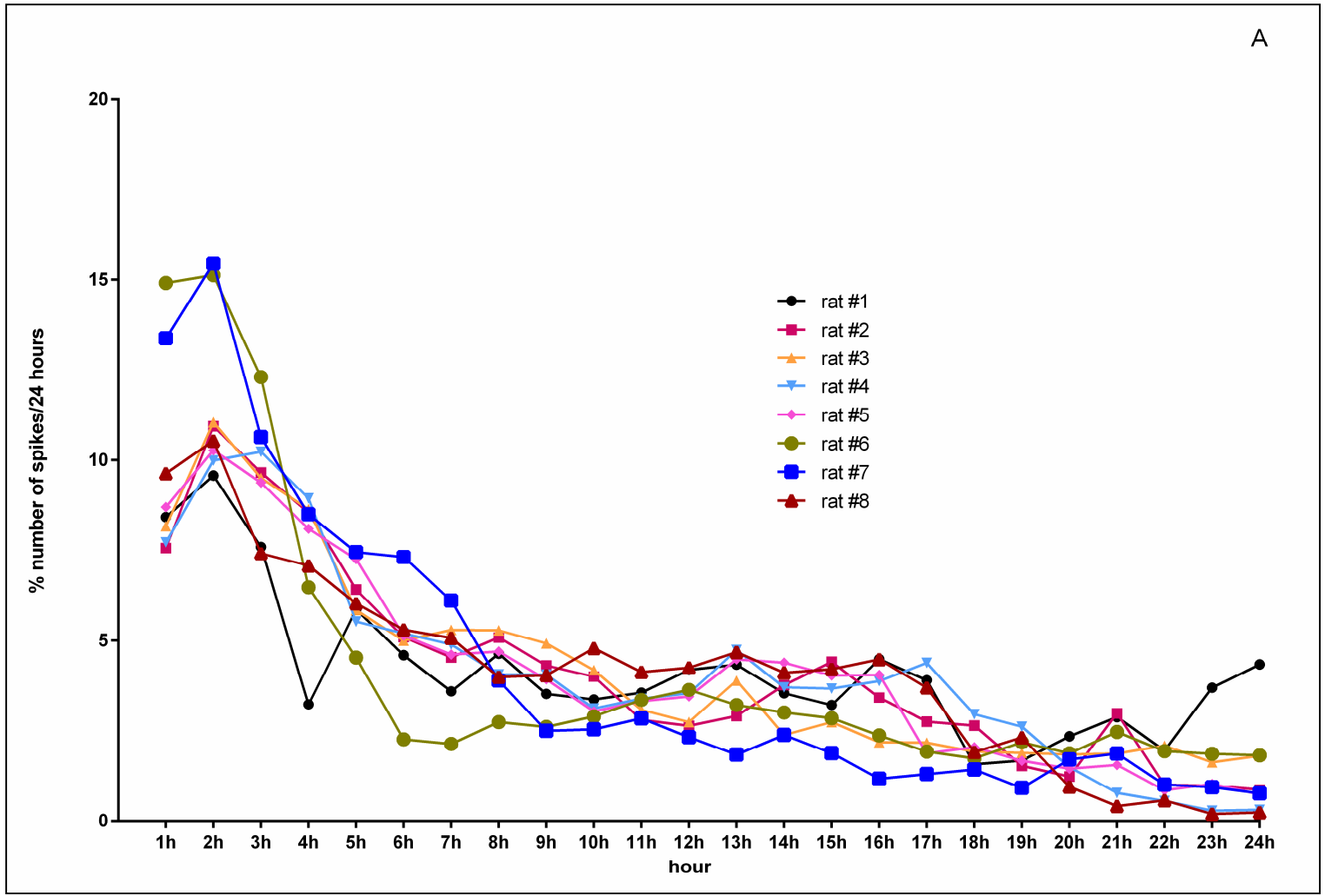

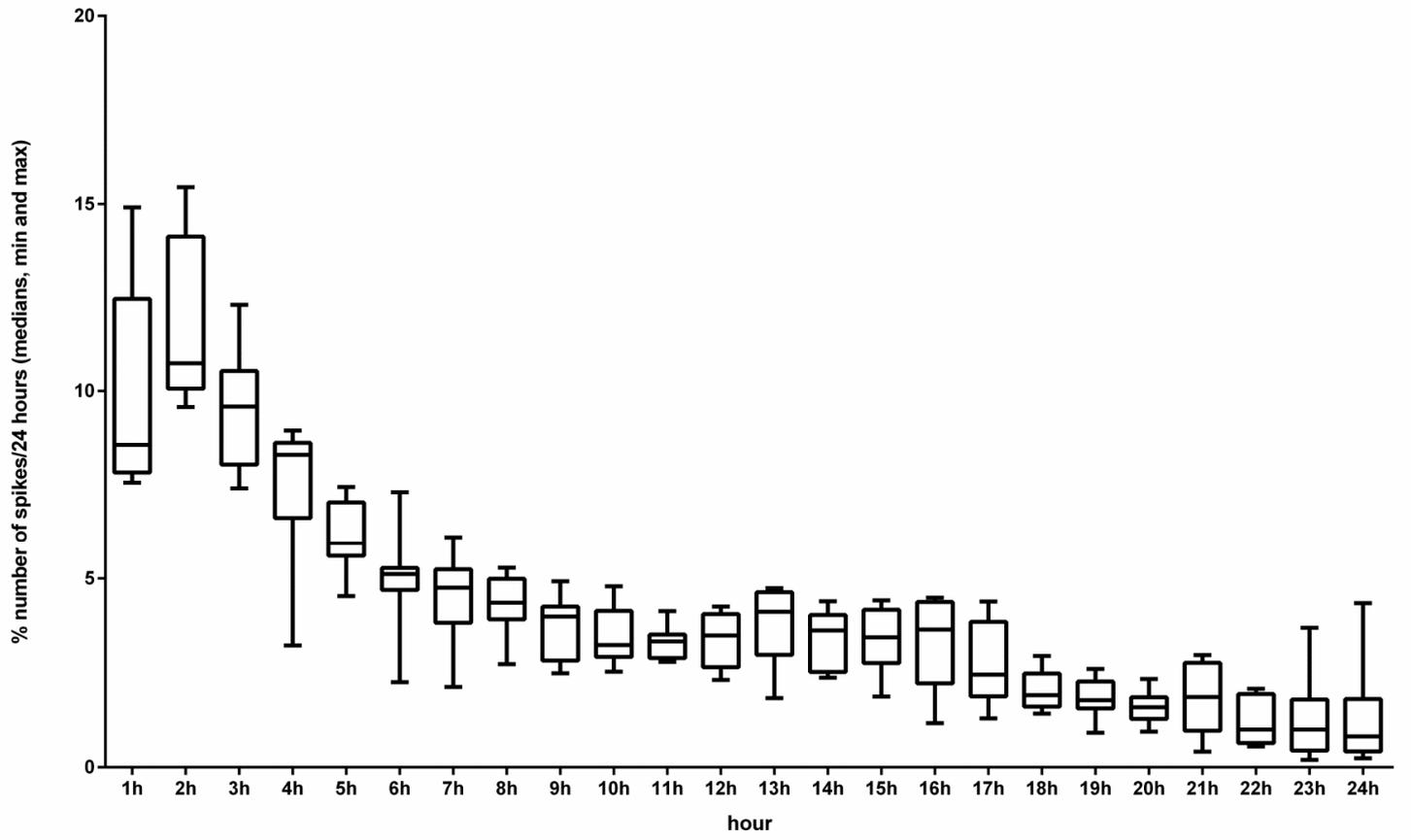

| | 1h | 2h | 3h | 4h | 5h | 6h | 7h | 8h | 9h | 10h | 11h | 12h | 13h | 14h | 15h | 16h | 17h | 18h | 19h | 20h | 21h | 22h | 23h | 24h |
|---|---|---|---|---|---|---|---|---|---|---|---|---|---|---|---|---|---|---|---|---|---|---|---|---|
| **rat #1** | u | u | p | u | u | u | u | u | ln | ln | ln | ln | ln | u | u | ln | ln | p | ln | ln | ln | p | ln | ln |
| p (bootstrap) | 0.00 | 0.00 | **0.23** | 0.00 | 0.02 | 0.00 | 0.00 | 0.00 | 0.00 | 0.01 | 0.00 | 0.00 | 0.00 | 0.00 | 0.00 | 0.00 | 0.00 | **0.91** | 0.00 | 0.00 | 0.00 | **0.99** | 0.01 | 0.00 |
| plausibility | | | *moderate* | | | | | | | | | | | | | | | *good* | | | | *likely* | | |
| **rat #2** | ln | ln | u | u | ln | ln | ln | ln | ln | ln | u | p | u | p | ln | ln | u | u | p | p | p | p | p | p |
| p (bootstrap) | 0.00 | 0.00 | 0.00 | 0.00 | 0.00 | 0.00 | 0.00 | 0.00 | 0.00 | 0.02 | 0.00 | **0.30** | 0.00 | **0.26** | 0.00 | 0.00 | 0.00 | 0.00 | **0.50** | **0.83** | **0.08** | **0.24** | **0.66** | **0.90** |
| plausibility | | | | | | | | | | | | *moderate* | | *moderate* | | | | | *moderate* | *moderate* | *likely* | *good* | *likely* | *moderate* |
| **rat #3** | u | p | ln | ln | ln | p | p | ln | ln | ln | ln | p | p | u | ln | p | u | p | p | p | ln | p | p | ln |
| p (bootstrap) | 0.03 | **0.26** | 0.01 | 0.00 | 0.00 | **0.26** | **0.23** | 0.00 | 0.00 | 0.00 | 0.00 | **0.30** | **0.42** | 0.06 | 0.02 | **0.47** | 0.06 | **0.29** | **0.12** | **0.62** | 0.00 | **0.16** | **0.11** | 0.00 |
| plausibility | | *moderate* | | | | *good* | *good* | | | | | *moderate* | *moderate* | | | *moderate* | | *moderate* | *good* | *good* | | *moderate* | *good* | |
| **rat #4** | u | p | u | u | ln | ln | ln | u | ln | u | u | u | u | ln | u | ln | ln | u | p | p | p | p | p | p |
| p (bootstrap) | 0.00 | **0.09** | 0.00 | 0.00 | 0.00 | 0.00 | 0.00 | 0.00 | 0.05 | 0.00 | 0.02 | 0.03 | 0.00 | 0.00 | 0.00 | 0.00 | 0.00 | 0.01 | **0.73** | **0.99** | **0.47** | **0.53** | **0.09** | **0.25** |
| plausibility | | *likely* | | | | | | | | | | | | | | | | | *moderate* | *moderate* | *likely* | *moderate* | *likely* | *moderate* |
| **rat #5** | p | p | ln | ln | ln | u | p | u | p | p | p | p | p | p | u | p | u | ln | p | p | p | u | p | p |
| p (bootstrap) | **0.11** | **0.24** | 0.00 | 0.00 | 0.00 | 0.01 | **0.18** | 0.00 | **0.50** | **0.41** | **0.10** | **0.94** | **0.18** | **0.14** | 0.00 | **0.64** | 0.44 | 0.01 | **0.25** | **0.19** | **0.36** | 0.01 | **0.48** | **0.09** |
| plausibility | *moderate* | *moderate* | | | | | *good* | | *good* | *moderate* | *likely* | *moderate* | *moderate* | *moderate* | | *moderate* | | | *moderate* | *moderate* | *moderate* | | *moderate* | *likely* |
| **rat #6** | u | p | ln | ln | ln | p | ln | ln | ln | ln | p | ln | ln | u | ln | ln | ln | ln | ln | u | ln | ln | p | ln |
| p (bootstrap) | 0.00 | **0.18** | 0.00 | 0.00 | 0.02 | **0.79** | 0.00 | 0.00 | 0.00 | 0.54 | **0.26** | 0.01 | 0.00 | 0.00 | 0.03 | 0.00 | 0.00 | 0.00 | 0.00 | 0.00 | 0.00 | 0.00 | **0.60** | 0.00 |
| plausibility | | *moderate* | | | | *good* | | | | | *moderate* | | | | | | | | | | | | *good* | |
| **rat #7** | u | u | ln | ln | ln | p | p | p | ln | ln | p | ln | ln | u | ln | p | ln | p | u | ln | ln | ln | ln | ln |
| p (bootstrap) | 0.00 | 0.05 | 0.00 | 0.00 | 0.00 | **0.82** | **0.70** | **0.92** | 0.00 | 0.00 | **0.12** | 0.00 | 0.00 | 0.00 | 0.00 | **0.66** | 0.01 | **0.42** | 0.01 | 0.00 | 0.00 | 0.00 | 0.00 | 0.00 |
| plausibility | | | | | | *good* | *good* | *moderate* | | | *moderate* | | | | | *likely* | | *likely* | | | | | | |
| **rat #8** | u | u | u | u | ln | p | p | ln | u | u | p | p | p | p | u | u | u | u | ln | p | p | ln | p | ln |
| p (bootstrap) | 0.00 | 0.01 | 0.03 | 0.02 | 0.00 | **0.36** | **0.09** | 0.00 | 0.01 | 0.00 | **0.91** | **0.35** | **0.11** | **0.66** | 0.04 | 0.00 | 0.00 | 0.00 | 0.00 | **0.86** | **0.28** | 0.00 | **0.68** | 0.00 |
| plausibility | | | | | | *moderate* | *likely* | | | | *good* | *moderate* | *good* | *good* | | | | | | *good* | *good* | | *good* | |

**Table 1**

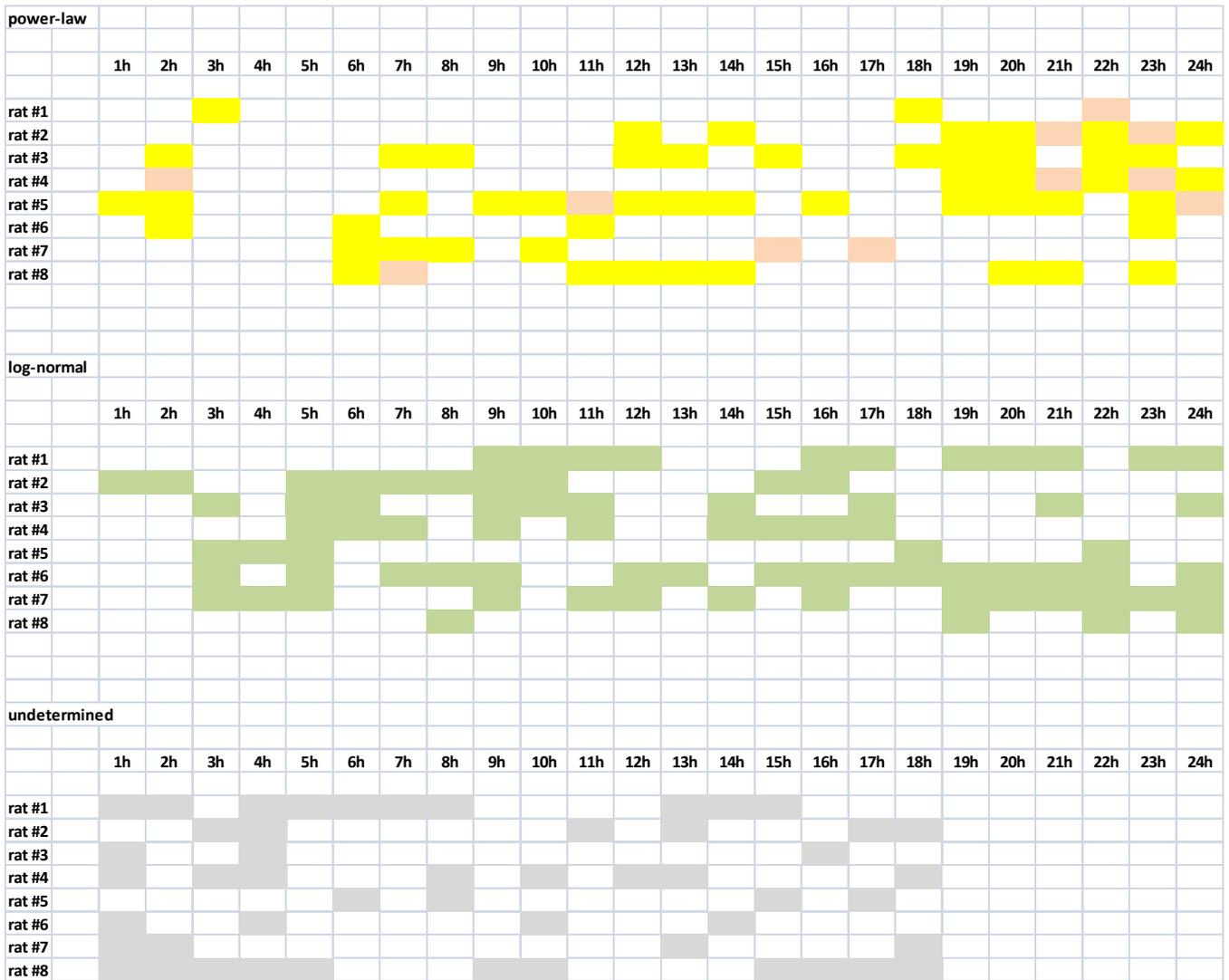

**Table 2**

**rat #1**

| | | 1h | 2h | 3h | 4h | 5h | 6h | 7h | 8h | 9h | 10h | 11h | 12h | 13h | 14h | 15h | 16h | 17h | 18h | 19h | 20h | 21h | 22h | 23h | 24h |
|---|---|---|---|---|---|---|---|---|---|---|---|---|---|---|---|---|---|---|---|---|---|---|---|---|---|
| log-normal | LR | 2.17 | -1.41 | -0.94 | -0.69 | -0.29 | 0.22 | 0.29 | 1.20 | -2.62 | -2.43 | -3.88 | -3.03 | -1.47 | 0.03 | -0.24 | -3.29 | -2.68 | 0.10 | -8.29 | -2.95 | -4.02 | 0.53 | -2.72 | -3.77 |
| | p | 0.0302 | 0.1580 | 0.3480 | 0.4930 | 0.7690 | 0.8240 | 0.7710 | 0.2290 | 0.0088 | 0.0051 | 0.0001 | 0.0025 | 0.1430 | 0.9730 | 0.8090 | 0.0010 | 0.0074 | 0.9230 | 0.0000 | 0.0032 | 0.0001 | 0.5940 | 0.0065 | 0.0002 |
| | | | | | | | | | | | | | | | | | | | | | | | | | |
| exponential | LR | 3.95 | 2.70 | 2.03 | 29.30 | 8.43 | 13.20 | 10.10 | 7.02 | 5.53 | 5.89 | 27.00 | 23.80 | 7.75 | 13.00 | 12.60 | 7.51 | 8.30 | 6.14 | 22.50 | 3.48 | 5.32 | 6.30 | 4.60 | 5.52 |
| | p | 0.0001 | 0.0069 | 0.0424 | 0.0000 | 0.0000 | 0.0000 | 0.0000 | 0.0000 | 0.0000 | 0.0000 | 0.0000 | 0.0000 | 0.0000 | 0.0000 | 0.0000 | 0.0000 | 0.0000 | 0.0000 | 0.0000 | 0.0005 | 0.0000 | 0.0000 | 0.0000 | 0.0000 |
| | | | | | | | | | | | | | | | | | | | | | | | | | |
| Poisson | LR | 3.01 | 10.60 | 5.62 | 19.80 | 3.59 | 7.69 | 10.70 | 6.96 | 10.20 | 7.80 | 20.70 | 18.80 | 10.80 | 7.24 | 10.60 | 7.44 | 10.30 | 3.97 | 25.60 | 9.59 | 12.90 | 3.75 | 12.10 | 11.40 |
| | p | 0.0026 | 0.0000 | 0.0000 | 0.0000 | 0.0003 | 0.0000 | 0.0000 | 0.0000 | 0.0000 | 0.0000 | 0.0000 | 0.0000 | 0.0000 | 0.0000 | 0.0000 | 0.0000 | 0.0000 | 0.0001 | 0.0000 | 0.0000 | 0.0000 | 0.0002 | 0.0000 | 0.0000 |
| | | | | | | | | | | | | | | | | | | | | | | | | | |
| | | | | | | | | | | | | | | | | | | | | | | | | | |
| | | | | | | | | | | | | | | | | | | | | | | | | | |
| p (bootstrap) | | 0.0000 | 0.0040 | 0.2322 | 0.0000 | 0.0210 | 0.0002 | 0.0000 | 0.0000 | 0.0006 | 0.0134 | 0.0000 | 0.0000 | 0.0026 | 0.0014 | 0.0000 | 0.0000 | 0.0000 | 0.9068 | 0.0000 | 0.0014 | 0.0000 | 0.9924 | 0.0110 | 0.0000 |
| | | | | | | | | | | | | | | | | | | | | | | | | | |
| favored distribution | | u | u | p | u | u | u | u | u | ln | ln | ln | ln | u | u | u | ln | ln | p | ln | ln | ln | p | ln | ln |
| plausibility for power-law | | | | moderate | | | | | | | | | | | | | | | good | | | | likely | | |
| alpha ± SD | | | | 3.82 ± 0.15 | | | | | | | | | | | | | | | 4.16 ± 1.05 | | | | 4.52 ± 0.78 | | |
| x_opt | | | | 797 | | | | | | | | | | | | | | | 111 | | | | 40 | | |

# rat #1

# Supplementary material

<table>
<thead>
<tr><th></th><th></th><th>1h</th><th>2h</th><th>3h</th><th>4h</th><th>5h</th><th>6h</th><th>7h</th><th>8h</th><th>9h</th><th>10h</th><th>11h</th><th>12h</th><th>13h</th><th>14h</th><th>15h</th><th>16h</th><th>17h</th><th>18h</th><th>19h</th><th>20h</th><th>21h</th><th>22h</th><th>23h</th><th>24h</th></tr>
</thead>
<tbody>
<tr><td>log-normal</td><td>LR</td><td>2.07</td><td>3.31</td><td>0.01</td><td>-6.98</td><td>-3.91</td><td>-17.30</td><td>-13.10</td><td>-6.32</td><td>-8.51</td><td>-13.30</td><td>-0.17</td><td>-0.44</td><td>-0.42</td><td>-0.01</td><td>-11.50</td><td>-22.20</td><td>3.51</td><td>0.58</td><td>-0.97</td><td>-0.69</td><td>-1.69</td><td>2.79</td><td>1.89</td><td>-1.15</td></tr>
<tr><td></td><td>p</td><td>0.0385</td><td>0.0009</td><td>0.9900</td><td>0.0000</td><td>0.0001</td><td>0.0000</td><td>0.0000</td><td>0.0000</td><td>0.0000</td><td>0.0000</td><td>0.8660</td><td>0.6590</td><td>0.6740</td><td>0.9900</td><td>0.0000</td><td>0.0000</td><td>0.0004</td><td>0.5600</td><td>0.3300</td><td>0.4880</td><td>0.0904</td><td>0.0053</td><td>0.0585</td><td>0.2490</td></tr>
<tr><td>exponential</td><td>LR</td><td>3.05</td><td>4.55</td><td>13.20</td><td>36.30</td><td>8.13</td><td>15.20</td><td>23.60</td><td>28.70</td><td>42.00</td><td>25.30</td><td>11.30</td><td>10.80</td><td>15.20</td><td>9.83</td><td>19.30</td><td>-5.93</td><td>39.90</td><td>37.30</td><td>8.81</td><td>5.65</td><td>5.32</td><td>9.49</td><td>8.67</td><td>6.11</td></tr>
<tr><td></td><td>p</td><td>0.0023</td><td>0.0000</td><td>0.0000</td><td>0.0000</td><td>0.0000</td><td>0.0000</td><td>0.0000</td><td>0.0000</td><td>0.0000</td><td>0.0000</td><td>0.0000</td><td>0.0000</td><td>0.0000</td><td>0.0000</td><td>0.0000</td><td>0.0000</td><td>0.0000</td><td>0.0000</td><td>0.0000</td><td>0.0000</td><td>0.0000</td><td>0.0000</td><td>0.0000</td><td>0.0000</td></tr>
<tr><td>Poisson</td><td>LR</td><td>2.17</td><td>4.29</td><td>17.00</td><td>34.90</td><td>7.19</td><td>72.00</td><td>83.80</td><td>17.20</td><td>63.90</td><td>65.50</td><td>5.64</td><td>4.62</td><td>5.37</td><td>5.80</td><td>53.70</td><td>71.00</td><td>6.46</td><td>3.65</td><td>4.44</td><td>5.97</td><td>6.56</td><td>5.10</td><td>3.52</td><td>5.22</td></tr>
<tr><td></td><td>p</td><td>0.0302</td><td>0.0000</td><td>0.0000</td><td>0.0000</td><td>0.0000</td><td>0.0000</td><td>0.0000</td><td>0.0000</td><td>0.0000</td><td>0.0000</td><td>0.0000</td><td>0.0000</td><td>0.0000</td><td>0.0000</td><td>0.0000</td><td>0.0000</td><td>0.0000</td><td>0.0003</td><td>0.0000</td><td>0.0000</td><td>0.0000</td><td>0.0000</td><td>0.0004</td><td>0.0000</td></tr>
<tr><td>p (bootstrap)</td><td></td><td>0.0000</td><td>0.0000</td><td>0.0000</td><td>0.0000</td><td>0.0000</td><td>0.0000</td><td>0.0000</td><td>0.0000</td><td>0.0000</td><td>0.0000</td><td>0.0238</td><td>0.3004</td><td>0.0000</td><td>0.2618</td><td>0.0000</td><td>0.0000</td><td>0.0000</td><td>0.0000</td><td>0.5044</td><td>0.8308</td><td>0.0832</td><td>0.2406</td><td>0.6560</td><td>0.8976</td></tr>
<tr><td>favored distribution</td><td></td><td>ln</td><td>ln</td><td>u</td><td>u</td><td>ln</td><td>ln</td><td>ln</td><td>ln</td><td>ln</td><td>ln</td><td>u</td><td>p</td><td>u</td><td>p</td><td>ln</td><td>ln</td><td>u</td><td>u</td><td>p</td><td>p</td><td>p</td><td>p</td><td>p</td><td>p</td></tr>
<tr><td>plausibility for power-law</td><td></td><td></td><td></td><td></td><td></td><td></td><td></td><td></td><td></td><td></td><td></td><td></td><td>moderate</td><td></td><td>moderate</td><td></td><td></td><td></td><td></td><td>moderate</td><td>moderate</td><td>likely</td><td>good</td><td>likely</td><td>moderate</td></tr>
<tr><td>alpha ±SD</td><td></td><td></td><td></td><td></td><td></td><td></td><td></td><td></td><td></td><td></td><td></td><td></td><td>3.45 ± 0.12</td><td></td><td>3.28 ± 0.14</td><td></td><td></td><td></td><td></td><td>4.15 ± 0.28</td><td>2.79 ± 0.31</td><td>3.00 ± 0.14</td><td>4.67 ± 0.51</td><td>4.65 ± 0.61</td><td>3.91 ± 0.28</td></tr>
<tr><td>x_tail</td><td></td><td></td><td></td><td></td><td></td><td></td><td></td><td></td><td></td><td></td><td></td><td></td><td>716</td><td></td><td>417</td><td></td><td></td><td></td><td></td><td>197</td><td>157</td><td>401</td><td>114</td><td>63</td><td>127</td></tr>
</tbody>
</table>

# rat #2

# Supplementary material

*rat #3*

| | | 1h | 2h | 3h | 4h | 5h | 6h | 7h | 8h | 9h | 10h | 11h | 12h | 13h | 14h | 15h | 16h | 17h | 18h | 19h | 20h | 21h | 22h | 23h | 24h |
|---|---|---|---|---|---|---|---|---|---|---|---|---|---|---|---|---|---|---|---|---|---|---|---|---|---|
| *log-normal* | LR | 1.39 | -0.53 | -3.12 | 12.90 | -5.66 | -12.80 | 0.86 | 4.53 | -15.60 | -15.50 | -22.70 | -0.17 | -1.09 | -6.29 | -0.99 | -1.81 | -2.65 | -1.15 | 0.15 | 3.34 | -3.20 | -0.11 | 13.50 | -2.16 |
| | p | 0.1640 | 0.5940 | 0.0018 | 0.0000 | 0.0000 | 0.0000 | 0.3900 | 0.0000 | 0.0000 | 0.0000 | 0.0000 | 0.8630 | 0.2780 | 0.0000 | 0.3230 | 0.0704 | 0.0080 | 0.2500 | 0.8850 | 0.0008 | 0.0014 | 0.9110 | 0.0000 | 0.0308 |
| *exponential* | LR | 2.05 | 2.53 | 1.63 | 16.80 | 32.70 | 20.20 | 0.60 | 3.25 | 10.60 | 9.21 | -1.73 | -6.87 | 11.80 | 4.91 | 10.10 | 11.00 | 12.50 | 9.29 | 12.70 | 10.30 | 13.60 | 11.90 | 11.10 | 11.40 |
| | p | 0.0399 | 0.0114 | 0.1040 | 0.0000 | 0.0000 | 0.0000 | 0.5500 | 0.0052 | 0.0000 | 0.0000 | 0.0838 | 0.0000 | 0.0000 | 0.0000 | 0.0000 | 0.0000 | 0.0000 | 0.0000 | 0.0000 | 0.0000 | 0.0000 | 0.0000 | 0.0000 | 0.0000 |
| *Poisson* | LR | 1.76 | 6.36 | 14.80 | 9.03 | 51.50 | 67.20 | 13.90 | 11.00 | 68.60 | 64.20 | 66.00 | 5.17 | 9.11 | 19.50 | 4.72 | 6.30 | 7.87 | 5.26 | 6.05 | 4.09 | 10.30 | 4.07 | 4.10 | 6.30 |
| | p | 0.0788 | 0.0000 | 0.0000 | 0.0000 | 0.0000 | 0.0000 | 0.0000 | 0.0000 | 0.0000 | 0.0000 | 0.0000 | 0.0000 | 0.0000 | 0.0000 | 0.0000 | 0.0000 | 0.0000 | 0.0000 | 0.0000 | 0.0000 | 0.0000 | 0.0000 | 0.0000 | 0.0000 |
| *p (bootstrap)* | | 0.0312 | 0.2576 | 0.0114 | 0.0000 | 0.0000 | 0.0000 | 0.2592 | 0.2266 | 0.0000 | 0.0000 | 0.0000 | 0.3004 | 0.4224 | 0.0000 | 0.4730 | 0.0556 | 0.0244 | 0.2888 | 0.1150 | 0.6180 | 0.0046 | 0.1550 | 0.1080 | 0.0026 |
| *favored distribution* | | u | p | ln | u | ln | ln | p | p | ln | ln | ln | p | p | ln | p | u | ln | p | p | p | ln | p | p | ln |
| *plausibility for power-low* | | | moderate | | | | | good | good | | | | moderate | moderate | | moderate | | | moderate | good | good | | moderate | good | |
| *alpha ± SD* | | | 3.82 ± 0.17 | | | | | 5.44 ± 0.23 | 3.69 ± 0.09 | | | | 4.49 ± 0.36 | 3.70 ± 0.14 | | 4.17 ± 0.45 | | | 4.12 ± 0.23 | 4.81 ± 0.24 | 4.81 ± 0.34 | | 4.72 ± 0.22 | 4.97 ± 0.21 | |
| *x_tot* | | | 450 | | | | | 463 | 932 | | | | 336 | 736 | | 255 | | | 212 | 166 | 121 | | 254 | 219 | |

**rat #3**

# Supplementary material

| | | 1h | 2h | 3h | 4h | 5h | 6h | 7h | 8h | 9h | 10h | 11h | 12h | 13h | 14h | 15h | 16h | 17h | 18h | 19h | 20h | 21h | 22h | 23h | 24h |
|---|---|---|---|---|---|---|---|---|---|---|---|---|---|---|---|---|---|---|---|---|---|---|---|---|---|
| log-normal | LR | 2.18 | -0.61 | 1.94 | 5.71 | -15.40 | -11.60 | -11.00 | -0.79 | -4.64 | 0.08 | -1.93 | -0.50 | -1.37 | -14.80 | -11.70 | -13.00 | -10.70 | -0.58 | -0.18 | -0.36 | 3.79 | -0.20 | -1.63 | -1.89 |
| | p | 0.0296 | 0.5400 | 0.0518 | 0.0000 | 0.0000 | 0.0000 | 0.0000 | 0.4300 | 0.0000 | 0.9350 | 0.0533 | 0.6200 | 0.1720 | 0.0000 | 0.0000 | 0.0000 | 0.0000 | 0.5620 | 0.8580 | 0.7170 | 0.0002 | 0.8400 | 0.1030 | 0.1360 |
| exponential | LR | 2.78 | 0.40 | 2.82 | 44.40 | 19.10 | 29.90 | 33.50 | 0.65 | -7.57 | 36.10 | 6.28 | 9.38 | 5.84 | 19.20 | 24.60 | 23.80 | 19.50 | 17.30 | 5.81 | 5.50 | 5.94 | 5.53 | 5.32 | 3.36 |
| | p | 0.0055 | 0.6910 | 0.0048 | 0.0000 | 0.0000 | 0.0000 | 0.0000 | 0.5180 | 0.0000 | 0.0000 | 0.0000 | 0.0000 | 0.0000 | 0.0000 | 0.0000 | 0.0000 | 0.0000 | 0.0000 | 0.0000 | 0.0000 | 0.0000 | 0.0000 | 0.0000 | 0.0008 |
| Poisson | LR | 2.61 | 3.34 | 2.21 | 38.40 | 77.60 | 63.00 | 64.20 | 16.70 | 37.60 | 3.89 | 9.54 | 8.38 | 6.01 | 68.20 | 53.50 | 52.00 | 46.50 | 8.63 | 5.90 | 5.15 | 2.91 | 4.55 | 6.46 | 6.63 |
| | p | 0.0092 | 0.0008 | 0.0271 | 0.0000 | 0.0000 | 0.0000 | 0.0000 | 0.0000 | 0.0000 | 0.0001 | 0.0000 | 0.0000 | 0.0000 | 0.0000 | 0.0000 | 0.0000 | 0.0000 | 0.0000 | 0.0000 | 0.0000 | 0.0037 | 0.0000 | 0.0000 | 0.0000 |
| p (bootstrap) | | 0.0006 | 0.0870 | 0.0000 | 0.0000 | 0.0000 | 0.0000 | 0.0000 | 0.0014 | 0.0000 | 0.0000 | 0.0452 | 0.0156 | 0.0340 | 0.0000 | 0.0000 | 0.0000 | 0.0000 | 0.0102 | 0.7306 | 0.9874 | 0.4680 | 0.5282 | 0.0908 | 0.2524 |
| favored distribution | | u | p | u | u | ln | ln | ln | u | ln | u | ln | u | u | ln | ln | ln | ln | u | p | p | p | p | p | p |
| plausibility for power-law | | | likely | | | | | | | | | | | | | | | | | moderate | moderate | likely | moderate | likely | moderate |
| alpha ± SD | | | 5.31 ± 0.35 | | | | | | | | | | | | | | | | | 3.21 ± 0.19 | 3.93 ± 0.42 | 4.44 ± 0.54 | 3.54 ± 0.25 | 3.62 ± 0.22 | 3.10 ± 0.22 |
| n_tail | | | 212 | | | | | | | | | | | | | | | | | 265 | 136 | 85 | 178 | 217 | 167 |

rat #4

**rat #5**

| | | 1h | 2h | 3h | 4h | 5h | 6h | 7h | 8h | 9h | 10h | 11h | 12h | 13h | 14h | 15h | 16h | 17h | 18h | 19h | 20h | 21h | 22h | 23h | 24h |
|---|---|---|---|---|---|---|---|---|---|---|---|---|---|---|---|---|---|---|---|---|---|---|---|---|---|
| log-normal | LR | -1.12 | -0.62 | -3.01 | -8.78 | -12.50 | -0.04 | 0.63 | -0.21 | 1.05 | -0.80 | -1.24 | -0.44 | -1.20 | -1.22 | -1.64 | -0.42 | -1.86 | -2.09 | -1.52 | -1.07 | -0.69 | -2.12 | -1.73 | -1.72 |
| | p | 0.2630 | 0.5390 | 0.0026 | 0.0000 | 0.0000 | 0.9690 | 0.6700 | 0.6380 | 0.2940 | 0.4250 | 0.2150 | 0.6580 | 0.2310 | 0.2220 | 0.1000 | 0.6720 | 0.0632 | 0.0366 | 0.1300 | 0.2830 | 0.4890 | 0.0339 | 0.0839 | 0.0848 |
| exponential | LR | 5.79 | 2.52 | 5.09 | 11.50 | 16.00 | 4.35 | 12.10 | 4.16 | 13.40 | 12.50 | 12.50 | 7.74 | 11.80 | 9.26 | 12.10 | 12.10 | 7.84 | 7.86 | 10.10 | 6.95 | 7.98 | 7.72 | 6.81 | 3.71 |
| | p | 0.0000 | 0.0118 | 0.0000 | 0.0000 | 0.0000 | 0.0000 | 0.0000 | 0.0000 | 0.0000 | 0.0000 | 0.0000 | 0.0000 | 0.0000 | 0.0000 | 0.0000 | 0.0000 | 0.0000 | 0.0000 | 0.0000 | 0.0000 | 0.0000 | 0.0000 | 0.0000 | 0.0002 |
| Poisson | LR | 8.84 | 7.00 | 18.60 | 41.90 | 65.20 | 6.99 | 5.88 | 9.31 | 4.70 | 7.60 | 4.95 | 5.46 | 7.71 | 7.52 | 7.52 | 6.81 | 8.14 | 8.43 | 5.96 | 4.85 | 3.64 | 6.13 | 7.13 | 6.66 |
| | p | 0.0000 | 0.0000 | 0.0000 | 0.0000 | 0.0000 | 0.0000 | 0.0000 | 0.0000 | 0.0000 | 0.0000 | 0.0000 | 0.0000 | 0.0000 | 0.0000 | 0.0000 | 0.0000 | 0.0000 | 0.0000 | 0.0000 | 0.0000 | 0.0003 | 0.0000 | 0.0000 | 0.0000 |
| p (bootstrap) | | 0.1148 | 0.2178 | 0.0000 | 0.0000 | 0.0000 | 0.0096 | 0.1814 | 0.0000 | 0.5006 | 0.4096 | 0.0988 | 0.9360 | 0.1844 | 0.1352 | 0.0044 | 0.6402 | 0.4356 | 0.0074 | 0.2490 | 0.1868 | 0.3564 | 0.0052 | 0.4806 | 0.0914 |
| favored distribution | | p | p | ln | ln | ln | u | p | u | p | p | p | p | p | p | u | p | u | ln | p | p | p | ln | p | p |
| plausibility for power-low | | moderate | moderate | | | | | good | | good | moderate | likely | moderate | moderate | moderate | | moderate | | | moderate | moderate | moderate | | moderate | likely |
| alpha ± SD | | 2.88 ± 0.06 | 4.23 ± 0.15 | | | | | 4.48 ± 0.24 | | 4.71 ± 0.23 | 3.73 ± 0.13 | 4.15 ± 0.18 | 4.04 ± 0.23 | 3.97 ± 0.42 | 3.81 ± 0.19 | | 3.81 ± 0.15 | | | 4.54 ± 0.29 | 3.61 ± 0.22 | 4.06 ± 0.25 | | 4.01 ± 0.39 | 2.99 ± 0.17 |
| n_tail | | 1554 | 734 | | | | | 319 | | 444 | 771 | 484 | 229 | 550 | 410 | | 659 | | | 160 | 268 | 184 | | 132 | 251 |

**rat #5**

**rat #6**

| | | 1h | 2h | 3h | 4h | 5h | 6h | 7h | 8h | 9h | 10h | 11h | 12h | 13h | 14h | 15h | 16h | 17h | 18h | 19h | 20h | 21h | 22h | 23h | 24h |
|---|---|---|---|---|---|---|---|---|---|---|---|---|---|---|---|---|---|---|---|---|---|---|---|---|---|
| *log-normal* | LLR | -1.67 | -0.85 | -6.16 | 31.90 | -2.43 | 2.88 | -3.30 | -5.11 | -6.22 | -1.99 | -0.31 | -2.12 | -4.83 | -1.91 | -4.68 | -4.98 | -6.61 | -6.07 | -2.74 | -6.79 | -2.55 | -3.76 | 0.36 | -2.57 |
| | *p* | 0.0940 | 0.3970 | 0.0000 | 0.0000 | 0.0150 | 0.0040 | 0.0010 | 0.0000 | 0.0000 | 0.0468 | 0.7600 | 0.0342 | 0.0000 | 0.0565 | 0.0000 | 0.0000 | 0.0000 | 0.0061 | 0.0000 | 0.0108 | 0.0002 | | 0.7220 | 0.0102 |
| | | | | | | | | | | | | | | | | | | | | | | | | | |
| *exponential* | LLR | 1.96 | -0.74 | 33.30 | 4.33 | 9.79 | 12.20 | 8.52 | 5.11 | 4.39 | 10.50 | 10.40 | 11.70 | 8.31 | 7.31 | 4.93 | 2.80 | -1.04 | -0.25 | 1.47 | 3.50 | 3.23 | 4.03 | 6.27 | 3.65 |
| | *p* | 0.0496 | 0.4610 | 0.0000 | 0.0000 | 0.0000 | 0.0000 | 0.0000 | 0.0000 | 0.0000 | 0.0000 | 0.0000 | 0.0000 | 0.0000 | 0.0000 | 0.0050 | 0.3010 | 0.8030 | 0.1400 | 0.0005 | 0.0012 | 0.0001 | | 0.0000 | 0.0003 |
| | | | | | | | | | | | | | | | | | | | | | | | | | |
| *Poisson* | LLR | 3.22 | 5.42 | 49.20 | 3.72 | 14.60 | 5.02 | 11.80 | 16.80 | 21.10 | 7.61 | 3.40 | 7.10 | 17.70 | 7.24 | 19.40 | 16.90 | 19.70 | 21.00 | 1.83 | 18.40 | 6.82 | 12.80 | 2.02 | -4.69 |
| | *p* | 0.0013 | 0.0000 | 0.0000 | 0.0002 | 0.0000 | 0.0000 | 0.0000 | 0.0000 | 0.0000 | 0.0000 | 0.0007 | 0.0000 | 0.0000 | 0.0000 | 0.0000 | 0.0000 | 0.0000 | 0.0000 | 0.0678 | 0.0000 | 0.0000 | 0.0000 | 0.0438 | 0.0000 |
| | | | | | | | | | | | | | | | | | | | | | | | | | |
| | | | | | | | | | | | | | | | | | | | | | | | | | |
| *p (bootstrap)* | | 0.0000 | 0.1804 | 0.0000 | 0.0000 | 0.0224 | 0.7942 | 0.0004 | 0.0000 | 0.0000 | 0.5416 | 0.2608 | 0.0094 | 0.0000 | 0.0276 | 0.0000 | 0.0000 | 0.0000 | 0.0000 | 0.0000 | 0.0000 | 0.0000 | 0.0028 | 0.0000 | 0.0026 | 0.0000 |
| | | | | | | | | | | | | | | | | | | | | | | | | | |
| *favored distribution* | | u | p | ln | u | p | ln | p | ln | ln | ln | u | p | ln | ln | u | ln | ln | ln | ln | ln | ln | ln | p | ln |
| *plausibility for power-law* | | | moderate | | | | good | | | | | moderate | | | | | | | | | | | | good | |
| *alpha ± SD* | | | 4.85 ± 0.29 | | | | 4.76 ± 0.82 | | | | | 4.43 ± 0.36 | | | | | | | | | | | | 3.83 ± 0.38 | |
| *n_tail* | | | 252 | | | | 113 | | | | | 242 | | | | | | | | | | | | 118 | |

**rat #6**

| rat #7 | | | | | | | | | | | | | | | | | | | | | | | | | | |
|---|---|---|---|---|---|---|---|---|---|---|---|---|---|---|---|---|---|---|---|---|---|---|---|---|---|---|
| | | | 1h | 2h | 3h | 4h | 5h | 6h | 7h | 8h | 9h | 10h | 11h | 12h | 13h | 14h | 15h | 16h | 17h | 18h | 19h | 20h | 21h | 22h | 23h | 24h |
| | | | | | | | | | | | | | | | | | | | | | | | | | | |
| log-normal | LR | | -1.85 | 0.35 | -6.35 | -5.45 | -7.73 | 0.15 | 4.26 | -0.43 | -4.12 | -0.36 | -3.43 | -4.77 | -1.81 | -4.01 | -0.30 | -4.25 | -1.52 | -1.66 | -2.88 | -3.97 | -2.09 | -3.80 | -3.08 | -2.86 |
| | p | | 0.0650 | 0.7230 | 0.0000 | 0.0000 | 0.0000 | 0.8780 | 0.0000 | 0.6650 | 0.0000 | 0.7230 | 0.0006 | 0.0000 | 0.0708 | 0.0001 | 0.7620 | 0.0000 | 0.1290 | 0.0967 | 0.0040 | 0.0001 | 0.0365 | 0.0001 | 0.0021 | 0.0043 |
| | | | | | | | | | | | | | | | | | | | | | | | | | | |
| exponential | LR | | 5.06 | 1.34 | 31.40 | 52.00 | 36.30 | 17.30 | 15.90 | 7.84 | 4.88 | 13.40 | 4.49 | 1.61 | 5.67 | 3.01 | 6.34 | -0.56 | 4.49 | 3.98 | 0.30 | 2.77 | 4.92 | 2.46 | 3.69 | 0.13 |
| | p | | 0.0000 | 0.1810 | 0.0000 | 0.0000 | 0.0000 | 0.0000 | 0.0000 | 0.0000 | 0.0000 | 0.0000 | 0.0000 | 0.1380 | 0.0000 | 0.0026 | 0.0000 | 0.5780 | 0.0000 | 0.0005 | 0.7630 | 0.0056 | 0.0000 | 0.0141 | 0.0002 | 0.8960 |
| | | | | | | | | | | | | | | | | | | | | | | | | | | |
| Poisson | LR | | 10.60 | 2.03 | 53.40 | 65.20 | 55.10 | 7.93 | 5.54 | 5.60 | 16.30 | 5.38 | 12.30 | 15.30 | 5.89 | 14.00 | 4.77 | 11.10 | 6.99 | 7.11 | 10.70 | 11.30 | 7.02 | 10.40 | 12.10 | 8.65 |
| | p | | 0.0000 | 0.0425 | 0.0000 | 0.0000 | 0.0000 | 0.0000 | 0.0000 | 0.0000 | 0.0000 | 0.0000 | 0.0000 | 0.0000 | 0.0000 | 0.0000 | 0.0000 | 0.0000 | 0.0000 | 0.0000 | 0.0000 | 0.0000 | 0.0000 | 0.0000 | 0.0000 | 0.0000 |
| | | | | | | | | | | | | | | | | | | | | | | | | | | |
| | | | | | | | | | | | | | | | | | | | | | | | | | | |
| p (bootstrap) | | | 0.0000 | 0.0490 | 0.0000 | 0.0000 | 0.0000 | 0.8162 | 0.6968 | 0.9158 | 0.0000 | 0.1172 | 0.0000 | 0.0000 | 0.0028 | 0.0000 | 0.6588 | 0.0000 | 0.4156 | 0.0106 | 0.0008 | 0.0000 | 0.0028 | 0.0000 | 0.0014 | 0.0042 |
| | | | | | | | | | | | | | | | | | | | | | | | | | | |
| favored distribution | | | u | u | ln | ln | ln | p | p | p | ln | p | ln | ln | u | ln | p | ln | p | u | ln | ln | ln | ln | ln | ln |
| plausibility for power-low | | | | | | | | good | good | moderate | | moderate | | | | | likely | | likely | | | | | | | |
| alpha ± SD | | | | | | | | 5.31 ± 0.20 | 5.16 ± 0.82 | 3.89 ± 0.17 | | 4.68 ± 0.84 | | | | | 4.34 ± 0.63 | | 3.70 ± 0.43 | | | | | | | |
| n_tail | | | | | | | | 300 | 302 | 242 | | 105 | | | | | 70 | | 84 | | | | | | | |

# Supplementary material

**rat #8**

| | | 1h | 2h | 3h | 4h | 5h | 6h | 7h | 8h | 9h | 10h | 11h | 12h | 13h | 14h | 15h | 16h | 17h | 18h | 19h | 20h | 21h | 22h | 23h | 24h |
|---|---|---|---|---|---|---|---|---|---|---|---|---|---|---|---|---|---|---|---|---|---|---|---|---|---|
| *log-normal* | LR# | -0.05 | -0.47 | 0.64 | 3.54 | 1.62 | -0.29 | 1.00 | -15.40 | 1.06 | 10.10 | 0.33 | -0.32 | 0.48 | 0.13 | -0.74 | 1.30 | 0.30 | -1.60 | -4.63 | 3.80 | 0.53 | -6.29 | 1.18 | -4.91 |
| | p | 0.9620 | 0.6400 | 0.5230 | 0.0004 | 0.3060 | 0.7750 | 0.3150 | 0.0000 | 0.2900 | 0.0000 | 0.7420 | 0.7490 | 0.6300 | 0.8990 | 0.4400 | 0.1930 | 0.7660 | 0.1090 | 0.0000 | 0.0001 | 0.5980 | 0.0000 | 0.2390 | 0.0000 |
| *exponential* | LR# | 4.29 | 3.62 | 5.44 | 2.03 | 6.06 | 12.50 | 11.00 | 12.70 | 2.58 | 2.74 | 11.30 | 11.90 | 15.40 | 11.20 | 2.27 | 11.40 | 9.56 | 36.10 | 9.80 | 9.03 | 16.40 | 20.50 | 9.82 | 16.20 |
| | p | 0.0000 | 0.0003 | 0.0000 | 0.0427 | 0.0000 | 0.0000 | 0.0000 | 0.0000 | 0.0100 | 0.0061 | 0.0000 | 0.0000 | 0.0000 | 0.0000 | 0.0233 | 0.0000 | 0.0000 | 0.0000 | 0.0000 | 0.0000 | 0.0000 | 0.0000 | 0.0000 | 0.0000 |
| *Poisson* | LR# | 5.59 | 5.93 | 7.24 | 4.62 | 7.06 | 8.62 | 3.29 | 29.00 | 3.61 | 4.30 | 4.60 | 3.59 | 5.44 | 4.49 | 5.98 | 1.99 | 10.40 | 25.60 | 17.10 | 1.55 | 5.86 | 23.20 | 2.65 | 20.10 |
| | p | 0.0000 | 0.0000 | 0.0000 | 0.0000 | 0.0000 | 0.0010 | 0.0000 | 0.0003 | 0.0000 | 0.0000 | 0.0003 | 0.0000 | 0.0000 | 0.0465 | 0.0000 | 0.1210 | 0.0000 | 0.0000 | 0.0000 | 0.0000 | 0.0080 | 0.0000 |  |  |
| *p (bootstrap)* | | 0.0002 | 0.0054 | 0.0304 | 0.0170 | 0.0000 | 0.3596 | 0.0888 | 0.0000 | 0.0000 | 0.0068 | 0.9146 | 0.3540 | 0.1138 | 0.6556 | 0.0374 | 0.0002 | 0.0000 | 0.0000 | 0.0000 | 0.8610 | 0.2762 | 0.0000 | 0.6758 | 0.0000 |
| *favored distribution* | | u | u | u | u | u | p | p | ln | u | u | p | p | p | p | u | u | u | u | ln | p | p | ln | p | ln |
| *plausibility for power-law* | | | | | | | moderate | likely | | | | good | moderate | good | good | | | | | | good | good | | good | |
| *alpha ± SD* | | | | | | | 3.93 ±0.15 | 3.55 ±0.14 | | | | 4.61 ±0.26 | 4.67 ±0.26 | 5.24 ±0.53 | 4.46 ±0.24 | | | | | | 4.40 ±1.30 | 4.47 ±0.21 | | 4.27 ±0.24 | |
| *n_tail* | | | | | | | 670 | 651 | | | | 339 | 281 | 171 | 342 | | | | | | 141 | 154 | | 123 | |

**rat #8**